# Homunculus: Auto-Generating Efficient Data-Plane ML Pipelines for Datacenter Networks

Tushar Swamy*, Annus Zulfiqar†, Luigi Nardi*‡,
Muhammad Shahbaz†, and Kunle Olukotun*
*Stanford University  †Purdue University  ‡Lund University

## ABSTRACT

Support for Machine Learning (ML) applications in networks has significantly improved over the last decade. The availability of public datasets and programmable switching fabrics (including low-level languages to program them) present a full-stack to the programmer for deploying in-network ML. However, the diversity of tools involved, coupled with complex optimization tasks of ML model design and hyperparameter tuning while complying with the network constraints (like throughput and latency), put the onus on the network operator to be an expert in ML, network design, and programmable hardware. This multi-faceted nature of in-network tools and expertise in ML and hardware is a road block for ML to become mainstream in networks, today.

We present Homunculus, a high-level framework that enables network operators to specify their ML requirements in a declarative, rather than imperative way. Homunculus takes as input, the training data and accompanying network constraints, and automatically generates and installs a suitable model onto the underlying switching hardware. It performs model design-space exploration, training, and platform code-generation as compiler stages, leaving network operators to focus on acquiring high-quality network data. Our evaluations on real-world ML applications show that Homunculus's generated models achieve up to 12% better F1 score compared to hand-tuned alternatives, while requiring only 30 lines of single-script code on average. We further demonstrate the performance of the generated models on emerging per-packet ML platforms to showcase its timely and practical significance.

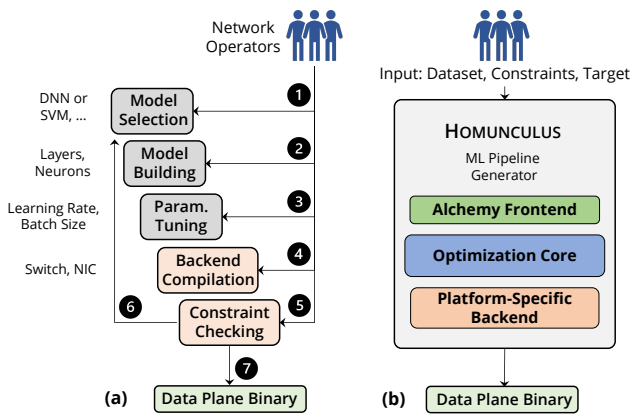

Figure 1: Comparison of data-plane ML pipeline development process. (a) Current approach: requires domain expertise in model selection, building, hyperparameter tuning, target compilation, and constraint checking; (b) Homunculus: requires only dataset, network performance constraints, and target type for resource estimation.

## 1 INTRODUCTION

It is the *reaction time* that dictates how robust, performant, and secure a given network is [100]. For example, (a) mitigating faults quickly (such as gray failures [39, 100]) minimizes network downtime and improves availability, (b) reacting to traffic imbalance due to short-lived traffic bursts lasting a few microseconds (microbursts) [101] swiftly alleviates network congestion [24, 33, 54, 91, 98] and server load [2, 49], (c) identifying malicious behavior early limits network disruptions and attacks [7, 16, 63, 87], and so on. Yet, at the same time, dealing with these events require complex operations, *e.g.,* Machine Learning (ML) inference, to decide what actions to perform as events happen; thereby, affecting our speed and response time to tackle such events [81, 85, 96].

Up until now, these inference operations were carried out on a logically-centralized control plane [63, 86, 91, 98], with decisions stored as flow rules in the network data plane (*i.e.,* switches and routers) [15]. The assumption, here, was that a decision made for the first packet will remain the same for all subsequent packets of a given flow (or connection). For routing and switching, where packets destined to a given destination must always reach the same server, this holds true. However, for performance and security objectives (like traffic engineering [7, 10, 19, 55] and threat mitigation [56]) which may require stateful processing, it is not the case, as network conditions can vary quickly within the duration of a given flow; hence, rendering any cached (per-flow) decisions stale and obsolete.

Recently, with the emergence of programmable switches (such as Intel Tofino [44, 45] with P4 programmable matchaction tables or MATs), SmartNICs and network accelerators (like Microsoft Catapult [75, 76], Azure AccelNet [29], Xilinx Alveo Data Center Accelerators [93] having a fieldprogrammable gate array or FPGA), the network data plane is no longer limited to flow-caching only. These data planes can now execute more complex operations and ML models directly in the network at line rate. For example, a P4-based switch can execute support-vector machines (SVMs) [27, 37, 63], K-Means [55], decision trees [4, 48] or binary neural networks (BNNs) [80, 81] directly in the data plane [96] to carry out tasks such as packet classification [48], load balancing [2, 49], resource-aware scheduling [60], and DDoS mitigation [52, 53, 56]. Similarly, SmartNICs and FPGA-based accelerators [18, 21, 29, 75, 76, 93] at end-hosts can run convolutional and recurrent deep neural networks (*i.e.,* CNNs and RNNs) to dynamically adjust congestion windows [5, 98], adapt bitrates for incoming video chunks [59, 97], predict network queue sizes [34] from the edge [32], and more. Emerging data plane platforms, like Taurus [85], have enabled per-packet ML inference for more complex ML algorithms, such as deep neural networks at line rate, as well.

Equipped with these programmable switches, NICs, and accelerators, datacenter networks can now react to network changes intelligently (running sophisticated ML models) and swiftly (at line speed). However, programming these data-plane ML pipelines today is extremely challenging, even for the most expert network operators in industry. There is a stark difference between designing a new feature (*e.g.,* adding or removing a protocol header) versus running a sophisticated model on a data plane [85, 96]. As shown in Figure 1, the latter requires domain expertise in ML model selection ❶, architecture construction, training ❷, and hyper-parameter optimization (HPO) ❸ as well as intimate familiarity with P4 and hardware-description languages (such as Verilog and VHDL) and emerging high-level DSLs (*e.g.,* Chisel [6] and Spatial [51]) ❹, while meeting network constraints (latency and line rate) ❺. This process is iterated by a network operator many times ❻ to compile these models on the underlying hardware backends ❼.

So, to build and execute these data-plane ML pipelines, a network operator would have to be an expert in ML model development, hardware architecture, and compiler design, all at the same time—a Sisyphean task for the network operators. Hence, despite the benefits, this steep barrier-to-entry for network operators has hindered the broader adoption of data-plane ML pipelines in datacenter networks to date.

If we recall, the aim behind *programmability* was not to modify the behavior of a data-plane device only, but to do so quickly and with ease [14, 61]. The current data-plane abstractions (like P4 [14]) make it easier for network operators to add new protocols and packet headers, without having to worry about the implementation details of underlying target backends. However, they are a poor fit for expressing complicated ML models; placing the onus on the network operators to explore, build, tune, and write these models in P4 (or similar low-level data-plane interfaces). Though, one can automate the process of generating the data-plane code for these models [100], the level of abstraction exposed to the network operators still requires them to manually and iteratively handle the various stages of model development (Figure 1a).

In this paper, keeping up with the original goal of programmability, we present Homunculus, a framework that alleviates the burden from network operators and automatically generates efficient data-plane ML pipelines for the various use cases (Figure 1b). With Homunculus, network operators now only have to specify (a) the training data set (*e.g.,* [87, 96]), (b) constraints of the operating environment, *i.e.,* minimum throughput and acceptable latency, and (c) particular targets for the ML pipeline to run on (*e.g.,* Tofino, Taurus [85], or else). Homunculus then performs design space exploration (DSE) [68] to find a model suitable for the given use case, tunes the hyper-parameters, and auto-generates the code for the specified backend. It iterates over these steps until the final output meets the constraints (or no feasible solution exists).

Homunculus leverages several key characteristics and recent advances in the fields of machine learning (including hyperparameter optimization [13, 67, 68, 84]) and networking. First, the recent proliferation of ML hardware accelerators (such as GPUs, TPUs, and NPUs) makes it possible to build large-scale systems for model search [36]; with open-source AutoML tools [12, 47], developers can now automate all stages of the ML development life-cycle, including model architecture search and training. Second, multi-objective black-box optimization frameworks (like Hyper-Mapper [68]) decouple model training from model search (for resource estimation and compliance)—demonstrating real-world successes in systems optimization, including hardware-software co-design [13], FPGA design [25, 51], computer vision [13, 67], robotics [79], and automated ML [84]. Homunculus augments the Keras [20] ML framework, which provides the necessary abstractions needed to build ML models, and incorporates hyperparameter suggestions from HyperMapper [68] to efficiently train these models. Third, Homunculus exploits the unique characteristics of datacenter networks (including operating constraints of minimum throughout and maximum acceptable latency) to further minimize the search space of data-plane ML models, by ruling out infeasible models during the search. Lastly, modern data centers—equipped with programmable data planes (Tofino), SmartNICs [42, 69], and FPGA-based network accelerators [29, 93] with open interfaces (such as P4, Spatial, and Verilog)—allow Homunculus to automatically generate efficient code for these individual backend targets. Emerging architectures, such as Taurus [85], further employ reconfigurable accelerators to increase the on-chip compute density for ML algorithms and are ripe for code automation and generation.

We summarize our key contributions as follows:

- **Alchemy DSL & Frontend:** We introduce Alchemy, an embedded DSL with a frontend for Homunculus that expresses user intent for a data-plane program in the form of objectives, data, and contsraints.
- **Optimization Core:** We illustrate methodologies for mapping multiple applications to a data-plane device. These methodologies form the optimization core of Homunculus, which takes the Alchemy input and explores the design space of ML model topologies, trains the models, and tests for constraints until a compliant and well-performing model is found.
- **Backend Generator:** We develop the Homunculus backend, which can generate Spatial [51] and P4 [14] code for the Taurus [85] or MAT-based [15] switches, respectively, using the ML model generated by the optimization core.
- **Empirical Evaluation:** We provide extensive evaluations and microbenchmarks of Homunculus using real-world applications (*e.g.*, anomaly detection, traffic classification, and botnet detection). Homunculus's generated ML pipelines achieve much higher F1 scores compared to hand-tuned baselines for Taurus and P4-SDNet FPGA [41, 85] backends: 83.1 and 68.75 versus 71.1 and 61.04, respectively. Moreover, for a botnet-detection model, originally hand-tuned for per-flow inference [9], Homunculus is able to search a per-packet model achieving an F1 score of 86.5—allowing networks to start detecting botnets immediately (at line speed), rather than waiting for the entire flow before taking any action.

## 2 BACKGROUND AND MOTIVATION

*ML for Networking.* ML algorithms are actively replacing existing heuristics (*e.g.*, for load balancing, packet classification, resource scheduling, and intrusion detection) in datacenter networks [16, 46, 58, 89, 99]. These heuristics are narrow in scope that tackle specific aspects of networking (*e.g.*, load balancing tailored to a leaf-spine topology [2]) or require manual tuning to adapt to new operating conditions [3, 54, 103]; therefore, when running with other heuristics, they typically result in sub-optimal behavior [16]. Furthermore, one-size-fit-all heuristics, which simultaneously handle multiple aspects of a network, are complex or even impossible for human operators to design [65, 83].

ML, on the other hand, can handle these complex relationships, efficiently. By training on the dataset captured for a given datacenter network, the model can customize itself to the particular environment. Over time, it can sample more data and retain itself to reflect changes in the network [16, 46, 58, 59, 97, 99]—understanding relationships between activities that network operators may not have been aware of. For example, instead of just matching incoming flows against a statically known set of IP addresses, ML can learn the correlation between fine-grain features (*e.g.*, connection duration, bytes transferred, protocol type, service type, packet size, and arrival time) to make informed decisions in new and unseen scenarios [23, 85, 87]. The rise in the number of ML-based use cases in last couple of years (*e.g.*, traffic classification [96], optimal routing [73], queue management [32, 34], and other network operations [16]) further highlights (and strengthen) the importance of ML for networking.

*Platforms for Data Plane ML.* With the advent of programmable data planes, the networking community is working actively on devising methods and platforms to run ML models on such data planes [80, 81, 85, 96]. Recent efforts primarily focus on transforming a manually-designed, pre-trained model on the underlying data-plane target using programming abstractions, like P4. For example, IIsy [96] is a data-plane pipeline that maps classical ML algorithms (such as SVM, KMeans, and decision trees) onto existing MATs in PISA switches. They exploit the structural similarity between these algorithms and the layout of MATs, which lend themselves to efficient hardware implementations. Still, devising hand-tuned algorithms to reach full model accuracy under strict network constraints (latency and throughout) and limited compute and storage resources of programmable switches is a challenge. For example, in their preliminary results, IIsy shows that an SVM consumes 8 MATs—25% of switch tables [44]—while KMeans requires 2 tables when compiled onto NetFPGA [57, 104] using Xilinx's P4-SDNet compiler [41]. Similarly, N2Net [81] provides a framework for compiling Binary Neural Networks (BNNs) to MATs. They transform existing neural network models as binary networks, by truncating model weights to a single bit value. Doing so impacts achievable model accuracy; but, the models can now run at line speed. The onus of developing a model such that it can meet the constraints and fit within the switch resources is still on the network operators—a single layer of a manually designed anomaly-detection DNN in N2Net takes up to 12 MATs [86]. Lastly, Taurus [85] introduces a new compute block (MapReduce) and an accompanying abstraction (Spatial [51]) in PISA switches—a dense array of compute and memory units, structured in a SIMD-like pattern, capable of executing various ML models. The SIMD patterns (map and reduce) enable efficient implementation of linear-algebra-based ML algorithms at line rate. However, like others above, the challenge again is on the network operator to manually design, transform, and program these models using low-level switch abstractions (*e.g.*, P4 and Spatial).

*Automated ML Frameworks.* ML methods are sensitive to a plethora of design decisions during model development, posing significant barriers to training and building custom ML models. This is particularly visible in the emerging field of artificial neural networks (ANNs), where ML developers are charged with selecting the right neural architectures, training procedures, and regularization methods along with tuning their hyperparameters to achieve high prediction accuracy [8, 65]. Today, the developers are left with tedious episodes of trial-and-error until they identify feasible set of choices for a particular ML application, given its dataset.

The systems and ML communities are placing considerable efforts in automating these decisions by providing useful abstractions for the developers [40, 68]. With open-source frameworks, such as AutoPytorch [105] and Auto-sklearn [28], a new field of automated ML (AutoML) [40] is emerging, with the goal of providing dutiful abstractions and runtimes to automatically generate trained ML models given only the training dataset as input. These algorithms cover a wide range of algorithms to choose from. For example, Neural Architecture Search (NAS) frameworks [12, 28, 105] for Convolutional Neural Networks (CNNs) attempt various permutations of well-tested, pre-build NN basic blocks (combinations of convolution kernels, activation functions, batch normalization, and more) to explore the design space of CNNs to find an optimal model for the task at hand. Similarly, framework like AutoKeras [47] let users specify ranges of tunable parameters (*e.g.,* whether to perform data augmentation and/or data normalization). The underlying systems explore this range, stochastically picking a set of (tunable) parameters while keeping track of various metrics (*e.g.,* accuracy or F1 score) for each iteration, ultimately selecting the best performing model. Lastly, frameworks such as Hyper-Mapper [51, 68] use advanced Bayesian optimization methods to formulate the task of model search as a black-box optimization problem with limited optimization budget. The goal is to establish algorithms that traverse the large search space with as little trial and error as possible [13, 67, 79].

In this paper, we extend upon these algorithms and devise an AutoML approach for generating data-plane ML pipelines to systematically and automatically search a design space and train models while adhering to the various performance constraints dictated by the datacenter network as well as target data-plane resources.

## 3 THE HOMUNCULUS DESIGN

HOMUNCULUS provides a high-level and declarative interface for network operators to specify their application objectives (*e.g.,* minimizing false positives in an anomaly-detection model or maximizing throughput of a traffic-classification algorithm). Network operators only specify the datasets—which can be proprietary or publicly available—along with the application objectives and a target backend to use (Figure 2). Given these datasets and objectives, HOMUNCULUS explores appropriate model architectures (*e.g.,* KMeans, SVMs, or DNNs) and automatically generates an optimal binary for the target backend, while respecting the environmental constraints (*i.e.,* network topology, hardware resources, and service-level objectives). It hides the low-level implementation details—that would otherwise require domain expertise in model selection and building, parameter tuning, backend compilation, and constraint checking (Figure 1a)—and allow operators to focus on curating network datasets and specifying performance goals (which they have expertise in).

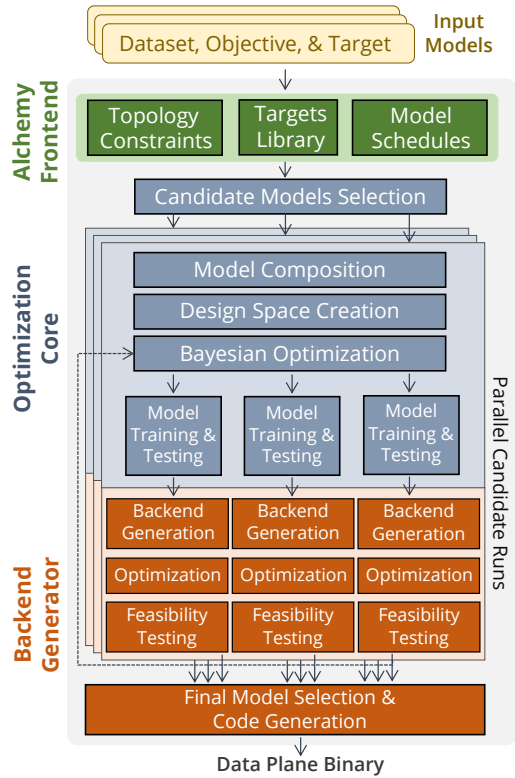

**Figure 2: High-level HOMUNCULUS framework with the Alchemy frontend (top), the optimization core (middle), and the backend generator (bottom).**

***Challenge: User objectives versus data-plane resources.***
HOMUNCULUS needs to carefully balance between meeting application objectives versus the available network and data-plane resources. Ultimately, it is the resources (and capabilities) of the data-plane target that determines what models are feasible. Certain models (and algorithms) may provide better performance with additional resources; the most efficient model will use as many resources as needed without over-provisioning. Similarly, moving to a new backend target (with different capabilities) may require reinterpretation of the objectives, possibly resulting in a different model.

By maintaining a repository of the resources and capabilities of the supported backends, HOMUNCULUS can automatically identify and evaluate various ML models for different backend configurations (Figure 2). For example, in a PISA switch, MATs are the limiting resource. Likewise, LUTs, BRAMs, and DSPs are the fundamental resources in an FPGA, which Homunculus can access from their respective hardware specifications or low-level compiler toolchains (*e.g.,* Barefoot P4 Studio [43] or Xilinx Vivado [92]).

The network topology can further guide Homunculus in accelerating the model search. For instance, a NIC typically operates in the 40–100Gbps range, whereas a ToR switch runs at terabits per second. Many model architectures can be eliminated by Homunculus as they may violate one or more of these requirements, effectively reducing the search space. As an example, a DNN that needs to operate at line rate, too many interations in the vector-matrix multiplication loop (for computing DNN layers) may bring down the device throughput and will, therefore, be eliminated by Homunculus during the exploration phase.

To achieve this balance between application objectives and resource constraints, the Homunculus framework is divided into three stages (Figure 2): the Alchemy DSL and frontend to solicit learned data-plane models by specifying datasets and objectives (§3.1); the optimization core for automated exploration, training, and testing of various ML algorithms and model architectures (§3.2); and the backend generator to transform and map candidate models and analyze and report target resource usage back to the optimization core or generate the final backend binary (§3.3).

*Running Example: An Anomaly-Detection Case Study.* Throughout the remainder of this section, we will use anomaly detection (AD) as a running example (Figure 3) to illustrate the various functions and capabilities of Homunculus's three stages on a recently proposed Taurus switch [85]. Based on the incoming traffic, the AD model marks packets as either benign or malicious. The model is trained using the NSL-KDD intrusion detection dataset [23], comprised of both benign traffic and different attack traces.

## 3.1 The Alchemy DSL & Frontend

We implement Alchemy as an embedded DSL (inside Python) that provides a declarative interface for the user to interact with the Homunculus framework. The user only specifies the dataset and objectives for their respective ML application (without explicitly writing the model definitions or doing model training). Alchemy, in essence, defines a *performance* abstraction that works alongside the functional description of the switch, provided by a network DSL (like P4).

Figure 3 shows our AD application written in the Alchemy DSL. Alchemy provides several constructs (classes and operators) to express and convey the programmer's intent to Homunculus, *i.e.,* the training dataset, application objectives and constraints, as well as interactions with other models or interfaces (to external devices). The DataLoader decorator wraps a custom function listing which datasets to use

```python
import homunculus
from homunculus.alchemy import
    DataLoader, Model, Platforms
import ad_loader

@DataLoader # training data loader definition
def wrapper_func():
    tnx, tny = ad_loader.load_from_file(
        "train_ad.csv")
    tsx, tsy = ad_loader.load_from_file(
        "test_ad.csv")
    return {
        "data": {"train": tnx, "test": tsx },
        "labels": {"train": tny, "test": tsy }}

# Specify the model of choice
model_spec = Model({
    "optimization_metric": ["f1"],
    "algorithm": ["dnn"],
    "name": "anomaly_detection",
    "data_loader": wrapper_func })

# Load platform
platform = Platforms.Taurus()
platform.constrain(
    "performance": {
        "throughput": 1, # GPkt/s
        "latency": 500 },  # ns
    "resources": { "rows": 16, "cols": 16 })

# Schedule model and generate code
platform.schedule(model_spec)
homunculus.generate(platform)
```

**Figure 3: Alchemy example syntax for the anomaly-detection use case. The pipeline is defined for the Taurus switch, scheduling a single model on the data plane (no model composition).**

in the model search. The function loads and preprocesses the dataset into a form that Homunculus can parse. For example, in Figure 3, the custom function, `ad_loader`, loads the NSL-KDD dataset (Line 7) and converts the categorical (multi-class) attacks into binary labels, *benign* and *malicious* (Line 8–14). Next, the Alchemy's Model class specifies the objective metric (and optional list of algorithms) to measure the performance of the candidate models. For example, we use the F1 score in our AD example (Line 17–18). To enforce environmental constraints (*e.g.,* data-plane resources, throughput, and latency), Alchemy supports the Platforms class that contains a list of supported backends (*e.g.,* Tofino or Taurus) with their accompanying constraints (Line 24–29). Moreover, Alchemy provides a collection of scheduling operators to specify how multiple models interact with each other (either sequentially or in parallel). Since, AD is the only application in our example, we schedule it by itself (Line 32). Finally, the generate function informs Homunculus to start model search and code generation.

### 3.1.1 Alchemy's Constructs: Classes and Operators.
Table 1 lists the various constructs available in Alchemy's embedded DSL.

| Construct | Symbol | Description |
|---|---|---|
| `Model` | Model(optimization_metric, algorithm, data_loader, ...) | Specify model objectives and datasets |
| `@DataLoader` | @DataLoader() | Load and preprocess model dataset |
| `Platforms` | Platforms.[Taurus, Tofino, FPGA] | Declare a backend target |
| `>, |` | Platforms.schedule(mdl1 > mdl2) or .schedule(mdl1 | mdl2) | Sequential > and parallel | composition |
| `IOMap` | IOMap(mapper_func) | Connects models' inputs and outputs |
| `@IOMapper` | @IOMapper([io_ins], [io_outs]) | Specifies input to output mapping |
| `<` | Platforms < (performance, resources) | Apply network and data-plane constraints |

Table 1: Alchemy's embedded DSL constructs: classes and operators.

- **Models** specify optimization objectives as well as a list of algorithms to guide the search space. If no algorithm is listed, Homunculus selects the best performing algorithm from among the entire list of supported algorithms. The model class also takes a decorated loader function to load and preprocess the labeled data.
- **Platforms** represents an instance of a physical device (like Taurus or Tofino), and informs Homunculus regarding the resources and configurations of the device (*e.g.,* MATs, LUTs, or BRAMs). It also lists the performance constraints (like latency and throughput). Homunculus's optimization phase (§3.2) considers all these constraints when selecting a model for the given target.
- **IOMap** describes how different components (models and platforms) connect with each other. It provides a mapping function that connects the inputs and outputs of these components and to the outside world.
- **Compositional operators** enable Homunculus to run multiple models on a given data plane, simultaneously. Models can either operate sequentially > or in parallel |, and can form a directed acyclic graph of any depth as long as the resources permit.

## 3.2 The Optimization Core

Homunculus's optimization core performs a design space search for model selection that maximizes the application objective while respecting the data-plane resources and network constraints. The core chooses from various ML algorithms and implements the model generation and data-plane mapping step as a Bayesian-optimization problem [22, 30, 72] when selecting the best performing configuration.

**3.2.1 Candidate Models Selection & Composition.** Homunculus aims to satisfy application objectives by exploring a variety of ML models (from the pool of supported algorithms). The core initiates multiple parallel runs to find the most efficient and performant model for the give application.[1] Some algorithms may consume less resources while others may perform better with bigger datasets. As a first step, the core tries to rule out as many algorithms as possible based on the data-plane platform and network constraints. For example, the number of multiplication and addition operations required for a modest DNN model can quickly exceed the computational capacity of a MAT-based switch; however, the mapping would indeed be feasible if ample MATs are available [81].

Next, when mapping multiple models to a given target, the optimization core ensures that (individual) model constraints are consistent with each other. For example, if one model operates at 1 GPkt/s throughput and feeds into another model operating at 0.5 GPkt/s, the first model must also operate at 0.5 GPkt/s. The core further disqualifies any such models that fail to meet these constraints.

**3.2.2 (Automated) Design Space Creation.** After creating a list of candidate algorithms, Homunculus's core uses the accompanying models' parameters and constraints to build a design space. Homunculus defines the search space by setting upper and lower bounds for these tunable parameters (*e.g.,* the minimum and maximum number of neurons in a DNN layer or the range of learning rate values attempted) to determine models' performance. These bounds are applied to three sets of variables:

- **Hyperparameters.** These are parameters of an ML model that are not decided by the training process. For a DNN, these includes parameters for the neural architecture search, such as the number of layers and neurons as well as training parameters (*e.g.,* learning rate and batch size). Traditionally, these parameters are hand tuned. They have a significant impact on the performance of the resulting model; however, the massive design space they cover makes hand-tuning extremely challenging. Homunculus instead explores the design space using Bayesian optimization (BO), by assigning upper and lower bounds to these parameters—typically calculated based on the target being considered.
- **Physical Resources.** The availability of resources on a data-plane platform has a large impact on the types of algorithms that the platform can support. For example, MATs in Tofino and the extent of loop unrolling in Taurus

---
[1] We take inspiration from other (iterative) compiler systems (such as Xilinx Vivado) that execute multiple parallel strategies to find the optimal resource placement and timing closure [92, 94]

place restrictions on the supported models. In Homunculus, we encode data-plane resources (such as CUs, MUs, and MATs) as feasibility constraints; if a certain algorithm configuration (*e.g.,* number of layers or neuron count of DNN) exceeds the provided resources, that configuration is marked infeasible and removed from the candidate set. Subsequent iterations of the Bayesian optimization will recommend model configurations that use less resources.

- ***Network Constraints.*** Network topology imposes further performance constraints. For example, a model running on a switch in the core of the network must operate at multi-terabits per second, whereas on an end-host NIC, it can run at 40–100Gbps. As with resources, Homunculus encodes these constraints as feasibility requirements, providing further opportunities for the optimization process to drop infeasible model configurations.

**3.2.3 Bayesian-Optimization (BO) Guided DSE.** With three classes of variables discussed earlier, Homunculus puts finite bound on a previously infinite design space. Each additional variable expands the design space and adds complexity to the optimization process. The difficulty of managing a space of this magnitude quickly outpaces human capabilities. And, at a certain point, it can even hamper automated search processes as well. Fortunately in Homunculus, rather than expanding the search, the additional variables for physical resources and network constraints help reduce the design space by disqualifying infeasible configurations, quickly. Even still, the search space is large enough to manage using basic heuristics, and demands efficient optimization processes to guide DSE.

***Formulating DSE as a black-box optimization problem.*** We now provide a formal definition of the optimization process and make it tractable by treating models and algorithms as opaque functions (*i.e.,* black boxes). In black-box optimization, the process is unaware of the internals of the models and, therefore, can swap different ML algorithms and configurations without changing the optimization method.

The black-box problem optimizes a (possibly noisy) function $f : \mathbb{X} \to \mathbb{R}$ over a domain of interest $\mathbb{X}$ that includes lower and upper bounds on the problem variables. The variables defining $\mathbb{X}$ can be real (continuous), integer, ordinal, or categorical as in [68]. The objective function $f$ for Homunculus is to maximize model performance (say F1 score) under the constraints ($c_i$) of network performance (latency, throughput) and resource consumption (such as LUTs, BRAM, CU, MU, MATs). $x^*$ is the optimal ML model configuration (say number of layers and neurons per layer for a DNN) that maximizes this objective while respecting the constraints.

Homunculus assumes that the function $f$ is in general expensive to evaluate, *e.g.,* it may take minutes or hours to evaluate one design $\mathbf{x} \in \mathbb{X}$ (even using a software model such as SARA [102] for Taurus), and that the derivatives of $f$ are in general not available. The second assumption implies that off-the-shelf gradient-based optimizers cannot be used to solve the optimization problem; the function is called *black box* because we cannot access other information than the output $y$ of the function $f$ (F1 score, resource utilization, latency/throughput) given an input value $\mathbf{x}$ (ML model configuration). Bayesian optimization (BO) achieves this by building a probabilistic surrogate model on $f$ based on the set of evaluated points. At each iteration, a new point is selected and evaluated using the surrogate model, and this model is updated to include the new point ($x_{t+1}, y_{t+1}$).

**3.2.4 Iterating over BO-Suggested Model Configurations.** In the optimization process, earlier, we treated the model evaluation (*e.g.,* checking accuracy, throughput, resource consumption) as a black box. In other words, the optimization process was unaware of the model training process as well as the environment required for testing feasibility and only saw inputs (ML parameter configurations) and outputs (model evaluation metrics such as F1 and its resource utilization).

The Keras ML framework is first delegated the responsibility of the training process. Since the user has specified a dataset and the optimization process has suggested hyperparameter configurations, Keras can train and test according to the metric designated by the application objectives. At this point, the only element left to evaluate are the feasibility constraints. To test this, Homunculus generates the hardware code (using ML model templates for each platform) required for mapping the model to the data plane (§3.3).

**3.2.5 Model Fusion.** To further save resources, some models can be combined into a single model. Models learning from similar datasets are most likely learning similar characteristics [88, 90]. Since most data coming into a data-plane model comes in the form of packets, data sets will have a number of features in common. Homunculus will assess the feature sets for similarities and if there are a certain number of features in common, it will attempt to build a single model to serve both datasets in the hope of sharing learned characteristics between tasks while also reducing resource usage by eliminating communication between models and also removing redundancies between learned weights.

## 3.3 The Backend Code Generator

Each target platform that Homunculus supports (*e.g.,* Taurus, Tofino, or FPGA) has an associated backend compiler to allow for performance and feasibility testing. Each backend is responsible for generating the platform's hardware code that can be tested in either simulation or on a physical testbed. We utilize cycle-accurate simulators (such as the SARA framework [102] for Taurus) that allow us to precisely measure resource utilization and latency/throughput of the ML model under test. Code is generated by assembling existing templates for common ML components before optimizing and testing.

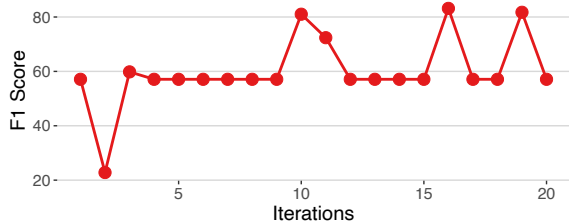

**Figure 4: Regret plot with F1 score metric for anomaly-detection DNN on Map-Reduce grid.**

After a predetermined number of optimization iterations, the best performing, constraint-compliant model is selected and its data-plane platform code and binary is generated. The amount of iterations an application takes to reach a desirable configuration is highly dependent on situational factors, such as the resources and data available. To give programmers an intuition of how this may affect their application, we show a plot in Figure 4 for the anomaly-detection application with the achieved F1 score in each optimization iteration. We see that while initial results are poor, HOMUNCULUS quickly begins to find a stable F1 score. Once it finds a variant that performs significantly better, it trades off between its earlier stable configuration and discovery of more high-performing variations effectively balancing exploitation of known parameter combinations and exploration of unknown ones.

***Template-based Code Generation for ML Models.*** To efficiently generate code for a backend, we use parameterized templates for commonly used operations. The parameters are calculated as a function of the optimization core's suggested configurations. These template blocks can then be assembled into larger blocks or even full packet pipelines (Figure 5). We consider the example of a Taurus switch, whose MapReduce block is programmed using the Spatial DSL [51]. We start with simple parameterized constructions, such as dot products, and build up to larger structures, like matrix multiplication and eventually DNNs. Simple dot products can be expressed as a map operation to perform element-wise multiplication with an addition-based reduction to combine the results into a scalar. These dot products can be nested inside an additional map operation to build a full DNN layer [74, 85]. The exact number of neurons and weights can be controlled by parameters dictated during the core optimization process, and the trained weights can be placed on the on-chip memory. These layers can then be stitched together by storing intermediate results for each layer in double-buffered SRAM blocks that feed into subsequent layers. Owing to the regular structure of ML algorithms (*e.g.,* KMeans and SVM) the code-generation process for other platforms (like Tofino using P4 templates) follow a similar methodology.

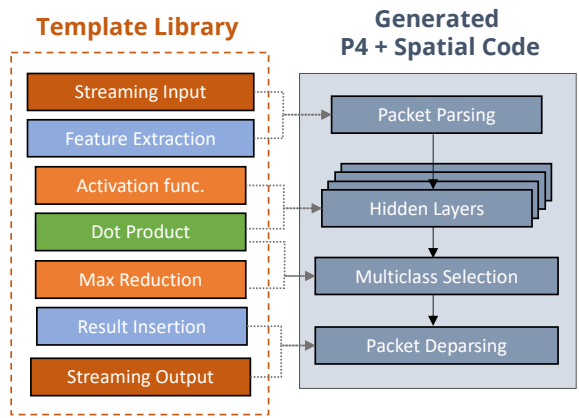

**Figure 5: Common ML operation templates are composed into larger building blocks and assemble into a complete data-plane ML pipeline.**

***Feasibility Constraint Testing for Generated Models.*** Once the hardware code has been generated for an ML model, it can be mapped into the testing infrastructure. The testing setup is required to provide verdicts on the feasibility constraints that the core optimization process is querying. In particular, the testing infrastructure is responsible for computing throughput and latency as well as identifying whether the application can be mapped within the available resources. This is done using hardware testbed platforms or cycle-accurate simulators (*e.g.,* Tungsten for Taurus or Xilinx Vivado for FPGAs) depending on the backend platform being used.

## 4 IMPLEMENTATION

***Alchemy DSL and Frontend.*** We implement the Alchemy frontend as a DSL embedded in Python. The embedded nature allows programmers to import HOMUNCULUS functions along with Alchemy constructs and use them in conjunction with existing Python primitives and libraries. The simplicity and versatility of Python make it an excellent base for our DSL, where these qualities are foundational to the design philosophy. In addition, the ubiquity of Python in machine learning, data science, and dataflow frameworks means that functions in these domains (such as Pandas [62], TensorFlow [1], or Theano [11]) can be incorporated into Alchemy's decorator functions (`IOMapper` or `DataLoader` or even its compositional operators.

***Optimization Core.*** HOMUNCULUS's optimization core employs Bayesian optimization to maximize user objectives while meeting feasibility constraints. To perform our optimization, we use HyperMapper [68], a framework for constrained multi-objective optimization along with standard machine-learning frameworks like Keras [20] and Tensorflow [1]. The design-space restrictions are parsed from the application's program (written in Alchemy) and formed into a JSON configuration file describing searchable parameters. This JSON file is fed to HyperMapper to start the optimization process. While the optimization is running, Homunculus receives regular parameter suggestions from HyperMapper, which are then evaluated for both the user's objective as well as adherence to feasibility constraints in target backends. The outcomes of these evaluations are then sent to HyperMapper to guide further design-space exploration.

*The Backend Code Generator.* For backend targets (like the Taurus [85] and FPGA), we use an intermediate hardware-description language, called Spatial [51]. Spatial relies on loop-level constructs to describe applications and translate them into a bitstream that can be applied to the underlying data-plane fabric (*e.g.,* Taurus or FPGA). The loop-level constructs make functional operators (like map and reduce) available, which allows our templates to be highly parallel and achieve high performance. In conjunction with Spatial, we use compiler frameworks (such as SARA [102]) to enable efficient mapping to tile-based reconfigurable architectures (like Taurus).

We also support common architectures such as the MAT-based pipelines found in switching architectures (*e.g.,* Tofino [45] or the P4-NetFPGA [41]). In these architectures, the number of available MATs becomes the constraining resource. We use IIsy [96] as a backend for mapping ML algorithms (such as SVMs or KMeans) to MATs. IIsy makes the relation between algorithm parameters and MATs explicit, a relation that can be exploited as a constraint by Homunculus. Homunculus will tune parameters in the algorithm that translate to MATs and attempts to fit generated models into the limited pipeline resources. For example, IIsy shows that an implementation of an SVM may use a MAT per feature. If the number of MATs is insufficient, Homunculus will try to remove less impactful features until the SVM model fits.

## 5 EVALUATION

In order to investigate the effectiveness of Homunculus, we test the compiler stack with real-world applications, addressing problems in network security and traffic classification. The models generated from Homunculus are constrained to 1 Gpkt/s line-rate throughput for all applications. We evaluate several microbenchmarks to demonstrate the capabilities of Homunculus framework in terms of resource budgeting and its adaptability to alternative switch architectures. For all these experiments, we setup HyperMapper to use the Random Forests [17] surrogate model, which is known to work well with systems workloads that require modeling of discrete parameters and non-continuous functions [68]. We select the Expected Improvement criterion [64] and a uniform random sampling initialization phase followed by Bayesian optimization iterations.

| Application | Features | # NN Param | F1 Score | CUs | MUs |
|---|---|---|---|---|---|
| Base-AD | 7 | 203 | 71.10 | 24 | 48 |
| Hom-AD | 7 | 254 | **83.10** | 41 | 67 |
| Base-TC | 7 | 275 | 61.04 | 31 | 59 |
| Hom-TC | 7 | 370 | **68.75** | 54 | 97 |
| Base-BD | 30 | 662 | 77.0 | 167 | 45 |
| Hom-BD | 30 | 501 | **79.8** | 53 | 151 |

**Table 2: Comparison of hand-tuned baseline models vs. Homunculus generated models.**

*Baseline Applications.* We evaluate the performance of Homunculus' generated models on the real-world anomaly-detection (AD) task. Our baseline is a hand-crafted AD model from [85] and [86], rewritten in Spatial [51] and trained offline on labeled packet-level traces from the NSL-KDD [23] dataset. We also evaluate Homunculus on the real-world traffic classification (TC) application, shown in IIsy [96], and botnet-detection (BD) application, found in Flowlens [9]. The TC application is built from IoT device traces in a data center and requires that an application correctly identifies the device type from packet-header features (packet size, Ethernet and IPv4 headers). The original TC models from IIsy [96] are statistical models (SVM, K-Means, Decision Trees), but we generate a hand-written DNN baseline with 3 hidden layers (10, 10, 5 neurons) to test against Homunculus's DNN generation capability for a fair comparison. The BD application is built from a dataset consisting of P2P applications that include traces from botnets (such as Storm and Waledac) as well as benign traces from uTorrent, Vuze, eMule, and Frostwire [77]. The botnet traffic can be segregated from benign P2P traffic by analyzing the histograms of packet sizes and packet inter-arrival time at the flow level.

### 5.1 Microbenchmarks

**5.1.1 Homunculus and Reaction Time.** The botnet detection (BD) problem has been studied, primarily, at the conversation level (tracking source and destination IP, while ignoring ports) [9, 66, 77]. The idea is to aggregate packet sizes and packet inter-arrival times into course-grained histograms (called *flowmarkers*) for up to 3,600 seconds (even on programmable switches like Tofino [9] using registers) before making a prediction of benign or malicious (botnet) flow. This is possible because botnets communicate via low-volume and high-duration flows compared to benign P2P applications, which makes them identifiable using their packet size and inter-arrival time histograms over the duration of their flows [9, 66, 77]. We demonstrate that by making predictions on a per-packet-level partial histograms, we can greatly reduce the reaction time of a botnet-detection model compared to using full flow-aggregated flowmarkers. Figure 6 shows histograms of packet sizes (bin size: 64 bytes) and inter-arrival times (bin size: 512 seconds) for benign and malicious classes averaged over the entire traffic. We observe that due to characteristic differences in the network traffic of benign and botnet P2P traffic, the resulting histograms of both kinds of applications start to look different—as certain bins are not expected to fill for botnet applications—early on with very little number of packets seen so far. This insight and empirical results serve as evidence for the need for per-packet ML and AutoML pipeline generators (such as HOMUNCULUS).

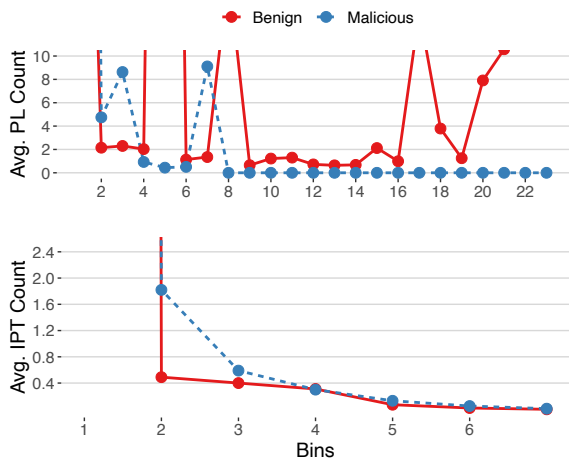

Figure 6: Botnet vs. Benign flow-level packet length (PL) and inter-arrival time (IPT) histograms averaged across all flows.

| Model | CUs | MUs |
| --- | --- | --- |
| DNN > DNN > DNN > DNN | 24 | 24 |
| DNN \| DNN \| DNN \| DNN | 24 | 24 |
| DNN > (DNN \| DNN) > DNN | 24 | 24 |

Table 3: Resource scaling for different application chaining strategies using a Taurus switch.

| Application | PCUs | PMUs |
| --- | --- | --- |
| AD: Part 1 | 44 | 81 |
| AD: Part 2 | 51 | 96 |
| AD: Fused | 48 | 83 |

Table 4: Fused Resource Usage

**5.1.2 HOMUNCULUS vs Baselines Resource Usage.** As shown in Table 2, for AD and TC applications, HOMUNCULUS is able to build models that outperform the hand-tuned baselines. This is because HOMUNCULUS is aware of the platform and environment where the application is being run and can make better use of available resources. Using more resources means choosing a bigger model (for example, more layers and neurons for a DNN) which directly implies an improvement in model's test performance. This is evident from the Compute Unit (CU) and Memory Unit (MU) usage numbers (in Table 2) on a Taurus platform. We argue that this is a point in favor for the compiler. If resources are unused, the platform will be under-utilized. However, it is challenging for users to cater models to the available resources without knowledge of the platform and environment. Bridging this gap results in potentially large increases in performance. In the case of AD, we see a full 12 point increase—this could easily be the difference between allowing malicious traffic to compromise a system and a stalwart defense.

For both the BD models, training was done on full flow-level histograms, while the F1 scores are reported on the per-packet-level partial histograms (120 Million test packets). The original FlowLens [9] model for BD used a flowmarker size of 151 bins (94 bins for packet size, the rest for inter-arrival time), while for our baseline and generated application, we use only 30 bins (23 for packet length and 7 for inter-arrival time) by fusing smaller bins into larger ones. This application is interesting because contrary to AD and TC, the baseline is a bigger model (4 hidden layers of 10 neurons each) in terms of parameter count and resource usage, as shown in Table 2. Yet, HOMUNCULUS manages to outperform the baseline model with a smaller model (10 hidden layers with smaller neuron count per layer) by distributing neurons across more layers. Notice that the baseline BD model is compute-intensive (using 167 CU units) due to larger hidden layer compute demands, while the generated model uses more memory units (151 MU) because it needs to store neurons and weights of a larger number of layers. While the FlowLens [9] BD model can perform with a perfect F1 score on large flow-level histograms accumulated over 3,600 seconds; despite our reduction in feature size and performing on per-packet-level histograms, we manage to get an F1 score of 77.0, which is much more reactive than waiting for an entire hour before labelling a malicious flow. Moreover, HOMUNCULUS manages to overcome the shortcomings of a hand-tuned design, and offers a better performing model with an F1 score of 79.8. Thus, not only are we able to reduce flowmarker size by 5× (hence increasing the number of flows we can handle on a switch proportionally), but we also reduce the reaction time from 3,600 seconds to a few hundred nanoseconds, while improving the F1 score.

**5.1.3 Multi-Application Scaling & Model Fusion.** We also demonstrate HOMUNCULUS's ability to support multiple applications on a single target. The Alchemy frontend allows a user to specify how different models interact with each other via sequential and parallel operators. Here, we show several examples of "app chaining" and show how it affects resource usage. Notation for chaining models in Table 3 follows our Alchemy operators. As a representative example, we chain copies of the anomaly-detection DNN in various configurations to emulate virtualization of user models on a single Taurus switch. In Table 3, we can see that the increase in resources for different chaining strategies stays constant with the number of models, regardless of the strategy itself. This is because additional logic for managing models is negligible and can be fitted into existing CUs, already in use; hence, allowing for much more efficient scaling of applications and sharing of backend resources.

For model fusion, in Table 4, we perform an experiment that divides the dataset of our AD application into two separate models. The two separate models map onto the same switch so they are each allocated half of the switch's resources. Since the datasets of the two models will share features, we can let Homunculus fuse them into a single model. Table 4 shows the resource counts for the two split models (AD: Part 1 and AD: Part 2) evaluated individually, and the resource count for a fused model that uses both datasets to create a single model. The resource count is about the same for the fused model since the two split models are each learning the same network characteristics. Instead of duplicating this knowledge, Homunculus encodes it into a single model, effectively cutting the resource usage by a factor of two.

## 5.2 End-to-End Hardware Evaluation

*Testbed Setup.* We use the Taurus testbed [85] for our end-to-end evaluation. A 32-port programmable Tofino Wedge100BF-32x switch running Stratum OS [71] is used to implement the PISA pipeline of the Taurus [85] switch. Its MAT processing pipelines are configured for pre- and post-processing, as necessary, to manage the Taurus ML core, which is emulated as a bump-in-the-wire. The switch bypasses its internal traffic through a Xilinx Alveo U250 FPGA [93] over a 100Gbps connection, which is used to emulate the MapReduce ML logic of a Taurus switch. A CMAC core [95] with an AXI interface is used to forward the packets to the FPGA. The control plane runs the ONOS controller [70] and a Python REST API is used to install forwarding rules on our switch. Two 80-core Intel Xeon servers generate and receive traffic via MoonGen [26]. Both, the baseline and the Homunculus-generated models are compiled to Verilog using the Spatial compiler and downloaded to the FPGA for evaluation. (All models operate at line-rate.)

### 5.2.1 Data-plane ML pipelines on a Taurus switch.

Table 5 summarizes the power and resource consumption of the seven models we test in hardware. Power consumption and resource utilization of loopback is due the *bump-in-the-wire* FPGA, which would be non-existent on an actual Taurus ASIC. We can observe from Table 5 that for the AD and TC applications, Homunculus generates larger models with higher layer/neuron count, which consume higher number of FPGA resources (Flip Flops, BRAM, LUTs) and, consequently, also consume higher power. Thus, there is a design trade-off here: if the better performing (higher F1 score) model from Homunculus is using higher number of parameters, that would translate to higher resource usage on the hardware (FPGA and ASIC alike) and, therefore, higher power consumption. The Homunculus model for the BD application uses slightly lower FPGA resources compared to the baseline model because the number of parameters in the baseline model is higher—resulting in larger difference in LUT consumption compared to other resources (LUTs store the parameters of a model in FPGA).

| Application | Model | LUT% | FFs% | BRAM% | Power (W) |
|---|---|---|---|---|---|
| Loopback | - | 5.36 | 3.64 | 4.15 | 15.131 |
| Base-AD | DNN | 6.55 | 4.30 | 4.15 | 16.969 |
| Hom-AD | DNN | 6.61 | 4.43 | 4.15 | 17.440 |
| Base-TC | DNN | 6.69 | 4.48 | 4.15 | 17.553 |
| Hom-TC | DNN | 7.48 | 4.77 | 4.15 | 18.405 |
| Base-BD | DNN | 7.29 | 4.68 | 4.15 | 17.807 |
| Hom-BD | DNN | 6.72 | 4.49 | 4.15 | 17.309 |

**Table 5: Resource consumption and utilization of applications running on the Taurus FPGA testbed [85].**

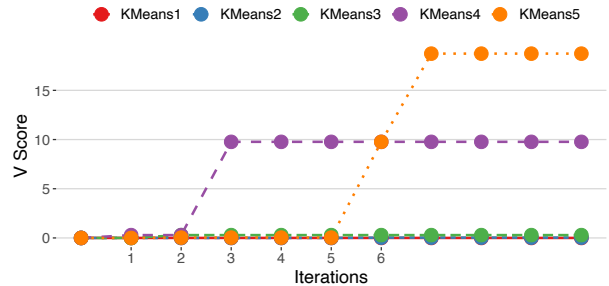

**Figure 7: Regret plot with V-Measure score metric for Kmeans on match-action tables (MATs).**

### 5.2.2 Data-plane ML pipelines on MAT-based switches.

Homunculus can support existing architectures, like a MAT-centered pipelines (*e.g.,* those found in Tofino-based switches). We show the results for an existing "ML for switches" platform, IIsy [96], plugged into Homunculus as a backend. Homunculus conforms the TC algorithm to the constraints given by IIsy and the switch hardware (MATs in this case). IIsy restricts a single MAT for each cluster, where a cluster corresponds to one of the five traffic classes. For switches with less tables, Homunculus creates more coarse-grain clusters, sacrificing fidelity in favor of resource usage.

In Figure 7, we show the V-measure score for a Homunculus-generated KMeans with varying resource availability. The objective is to cluster traffic from multiple IoT devices into a number of groups using packet header features. In the IIsy framework, a KMeans implementation can be mapped onto MATs. However, each cluster grouping takes up an additional MAT. In Figure 7, we see five generated implementations of KMeans traffic classification using the IIsy backend, each with different resource constraints. KMeans5 (K5) represents the application with 5 available tables, K4 has 4, and so on. Homunculus automatically generates models to fit each of the different resource constraints by dropping clusters and opting for coarser groupings at the cost of accuracy (as measured by the V-score).

## 6 RELATED WORK

***Automated Machine Learning.*** Machine learning (ML) has achieved considerable successes in recent years and an ever-growing number of disciplines rely on it. However, this success crucially relies on ML experts who are a scarce human resource. As the complexity of these tasks is often beyond non-ML-experts, the rapid growth of ML applications has created a demand for off-the-shelf ML methods that can be used easily and without expert knowledge. The resulting research area that targets progressive automation of machine learning is called Automated ML (or AutoML) [40]. AutoML has been a popular topic in the deep learning community. With so many hyperparameters and ever-increasing model sizes, tuning hyperparameters by hand is an extremely challenging task and developers have started relying on AutoML to explore the space of possible neural architectures and their hyperparameters. AutoML neural architecture strategies often revolve around arrangement of pre-built blocks rather than building the model from scratch [35, 78] to reduce the complexity of the optimization space, however limiting the versatility of resulting models.

AutoML tasks, typically do not share the unique and stringent set of constraints that arise in datacenter networking that ultimately allow Homunculus to handle, otherwise, untenable search spaces. Computer networks have several requirements that need to be taken into account jointly with the neural architecture search. Multi-objective optimization is a crucial matter because real-world applications often rely on a trade-off between several objectives [50, 68]. Derivatives are usually not available because of the discrete input variables and noisy black-box functions, or are impractical to compute and the feasibility of an experiment can not always be determined in advance [31]. These problems are particularly difficult when the feasible region is relatively small, and it may be prohibitive to even find a feasible experiment, let alone an optimal one. All of these features are common requirements in datacenter networking but rarely exposed in AutoML frameworks, which usually focus on accuracy of the ML model alone. In general, there have been only a few attempts to date at AutoML-driven systems in the networking space. An example is a traffic analysis application [38] where the authors employed an AutoML approach. This work focuses on refining network data representations and fits well with systems, such as Homunculus, that can consume and utilize this data.

***Code Generation for Data Plane.*** Recent works have explored code generation for data and control planes. Mantis [100] generates data plane and switch-local control plane software optimized for reactivity, allowing a faster control plane that can react to congestion conditions in 10s of $\mu$-seconds. Lucid [82] generates efficient P4 code for the data plane via a high-level language that allows to specify control-plane functionality, as well. These control-plane functions are mapped to the data plane, allowing faster control decisions compared to switch-local control plane. While these efforts are primarily meant for data-plane and control-plane code generation to enable either reactivity or data-plane control, our approach is specifically meant for generating efficient ML pipelines for data-plane platforms that fit within the available resources and perform within the given constraints.

***Configurable Hardware.*** Plasticine [74], a Coarse-Grained Reconfigurable Array (CGRA), provides a grid of compute and memory units, which can be reconfigured according to the dataflow graph of an application. This SIMD grid of compute and memory units provides the necessary flexibility to implement neural networks in the data plane that are not realizable in the VLIW MAT pipeline only. Taurus' MapReduce block is based on this architecture, which has been tailored to support streaming data on switches and NICs [85]. Reconfigurable Match Table (RMT) is an example of a reconfigurable architecture that has been specifically designed while considering the domain characteristics of the network data plane [15]. While the two platforms are structurally different (SIMD vs VLIW), Homunculus is agnostic to these architectural variations and only needs to query the resource utilization and performance metrics to generate platform-specific code for either of them.

## 7 CONCLUSION

We presented Homunculus, a compiler for building data-plane ML pipelines from high-level directives—giving network operators an easy way to inform the compiler of their needs rather than having to spend time tuning hyperparameters and trying to manage the various demands of the network data-planes and topology. It is clear that human operators are no longer the best choice for deploying algorithms in a datacenter network. The environment is far too complex for humans to manage. Instead, they should describe their intent and let automated systems handle the rest. With Homunculus, we move one step toward this era of declarative, intent-based networking.

# REFERENCES

[1] ABADI, M., BARHAM, P., CHEN, J., CHEN, Z., DAVIS, A., DEAN, J., DEVIN, M., GHEMAWAT, S., IRVING, G., ISARD, M., KUDLUR, M., LEVENBERG, J., MONGA, R., MOORE, S., MURRAY, D. G., STEINER, B., TUCKER, P., VASUDEVAN, V., WARDEN, P., WICKE, M., YU, Y., AND ZHENG, X. TensorFlow: A System for Large-Scale Machine Learning. In *USENIX OSDI* (2016).

[2] ALIZADEH, M., EDSALL, T., DHARMAPURIKAR, S., VAIDYANATHAN, R., CHU, K., FINGERHUT, A., LAM, V. T., MATUS, F., PAN, R., YADAV, N., AND VARGHESE, G. CONGA: Distributed Congestion-aware Load Balancing for Datacenters. In *ACM SIGCOMM* (2014).

[3] ALIZADEH, M., GREENBERG, A., MALTZ, D. A., PADHYE, J., PATEL, P., PRABHAKAR, B., SENGUPTA, S., AND SRIDHARAN, M. Data Center TCP (DCTCP). In *ACM SIGCOMM* (2010).

[4] AMOR, N. B., BENFERHAT, S., AND ELOUEDI, Z. Naive Bayes vs Decision Trees in Intrusion Detection Systems. In *ACM Symposium on Applied Computing (2004)* (2004).

[5] ARASHLOO, M. T., LAVROV, A., GHOBADI, M., REXFORD, J., WALKER, D., AND WENTZLAFF, D. Enabling Programmable Transport Protocols in High-Speed NICs. In *USENIX NSDI (2020)* (2020).

[6] BACHRACH, J., VO, H., RICHARDS, B., LEE, Y., WATERMAN, A., AVIŽIENIS, R., WAWRZYNEK, J., AND ASANOVIĆ, K. Chisel: Constructing Hardware in a Scala Embedded Language. In *DAC* (2012).

[7] BAKKER, J., NG, B., SEAH, W. K., AND PEKAR, A. Traffic Classification with Machine Learning in a Live Network. In *IFIP/IEEE Symposium on Integrated Network and Service Management (IM)* (2019).

[8] BALDI, P., AND SADOWSKI, P. J. Understanding Dropout. *Advances in Neural Information Processing Systems 26* (2013), 2814–2822.

[9] BARRADAS, D., SANTOS, N., RODRIGUES, L., SIGNORELLO, S., RAMOS, F. M. V., AND MADEIRA, A. FlowLens: Enabling Efficient Flow Classification for ML-based Network Security Applications. In *NDSS* (2021).

[10] BENSON, T., AKELLA, A., AND MALTZ, D. A. Network Traffic Characteristics of Data Centers in the Wild. In *ACM IMC* (2010).

[11] BERGSTRA, J., BREULEUX, O., BASTIEN, F., LAMBLIN, P., PASCANU, R., DESJARDINS, G., TURIAN, J., WARDE-FARLEY, D., AND BENGIO, Y. Theano: A CPU and GPU Math Expression Compiler. In *SciPy* (2010).

[12] BISONG, E. Google AutoML: Cloud Vision. In *Building Machine Learning and Deep Learning Models on Google Cloud Platform*. Springer, 2019, pp. 581–598.

[13] BODIN, B., NARDI, L., ZIA, M. Z., WAGSTAFF, H., SREEKAR SHENOY, G., EMANI, M., MAWER, J., KOTSELIDIS, C., NISBET, A., LUJAN, M., FRANKE, B., KELLY, P. H., AND O'BOYLE, M. Integrating Algorithmic Parameters into Benchmarking and Design Space Exploration in 3D Scene Understanding. In *ACM PACT* (2016).

[14] BOSSHART, P., DALY, D., GIBB, G., IZZARD, M., MCKEOWN, N., REXFORD, J., SCHLESINGER, C., TALAYCO, D., VAHDAT, A., VARGHESE, G., AND WALKER, D. P4: Programming Protocol-independent Packet Processors. *ACM SIGCOMM CCR* (2014).

[15] BOSSHART, P., GIBB, G., KIM, H.-S., VARGHESE, G., MCKEOWN, N., IZZARD, M., MUJICA, F., AND HOROWITZ, M. Forwarding Metamorphosis: Fast Programmable Match-Action Processing in Hardware for SDN. In *ACM SIGCOMM* (2013).

[16] BOUTABA, R., SALAHUDDIN, M. A., LIMAM, N., AYOUBI, S., SHAHRIAR, N., ESTRADA-SOLANO, F., AND CAICEDO, O. M. A Comprehensive Survey on Machine Learning for Networking: Evolution, Applications and Research Opportunities. *Journal of Internet Services and Applications* (2018).

[17] BREIMAN, L. Random Forests. *Machine learning 45*, 1 (2001), 5–32.

[18] CAULFIELD, A. M., CHUNG, E. S., PUTNAM, A., ANGEPAT, H., FOWERS, J., HASELMAN, M., HEIL, S., HUMPHREY, M., KAUR, P., KIM, J.-Y., LO, D., MASSENGILL, T., OVTCHAROV, K., PAPAMICHAEL, M., WOODS, L., LANKA, S., CHIOU, D., AND BURGER, D. A Cloud-scale Acceleration Architecture. In *IEEE MICRO* (2016).

[19] CHEN, L., LINGYS, J., CHEN, K., AND LIU, F. AuTO: Scaling Deep Reinforcement Learning for Datacenter-scale Automatic Traffic Optimization. In *ACM SIGCOMM* (2018).

[20] CHOLLET, F., ET AL. Keras: The Python Deep Learning Library. *Astrophysics Source Code Library* (2018).

[21] CHUNG, E., FOWERS, J., OVTCHAROV, K., PAPAMICHAEL, M., CAULFIELD, A., MASSENGILL, T., LIU, M., LO, D., ALKALAY, S., HASELMAN, M., ABEYDEERA, M., ADAMS, L., ANGEPAT, H., BOEHN, C., CHIOU, D., FIRESTEIN, O., FORIN, A., GATLIN, K. S., GHANDI, M., HEIL, S., HOLOHAN, K., EL HUSSEINI, A., JUHASZ, T., KAGI, K., KOVVURI, R. K., LANKA, S., VAN MEGEN, F., MUKHORTOV, D., PATEL, P., PEREZ, B., RAPSANG, A., REINHARDT, S., ROUHANI, B., SAPEK, A., SEERA, R., SHEKAR, S., SRIDHARAN, B., WEISZ, G., WOODS, L., YI XIAO, P., ZHANG, D., ZHAO, R., AND BURGER, D. Serving DNNs in Real Time at Datacenter Scale with Project Brainwave. *IEEE Micro* (2018).

[22] CURRIN, C., MITCHELL, T., MORRIS, M., AND YLVISAKER, D. A Bayesian Approach to the Design and Analysis of Computer Experiments. Tech. rep., Oak Ridge National Laboratory, TN (USA), 1988.

[23] DHANABAL, L., AND SHANTHARAJAH, S. A Study on NSL-KDD Dataset for Intrusion Detection System Based on Classification Algorithms. *International Journal of Advanced Research in Computer and Communication Engineering 4*, 6 (2015), 446–452.

[24] DONG, M., LI, Q., ZARCHY, D., GODFREY, P. B., AND SCHAPIRA, M. PCC: Re-Architecting Congestion Control for Consistent High Performance. In *USENIX NSDI* (2015).

[25] EJJEH, A., ADVE, V., AND RUTENBAR, R. A. Studying the Potential of Automatic Optimizations in the Intel FPGA SDK for OpenCL. In *ACM/SIGDA FPGA* (2020).

[26] EMMERICH, P., GALLENMÜLLER, S., RAUMER, D., WOHLFART, F., AND CARLE, G. Moongen: A Scriptable High-speed Packet Generator. In *ACM IMC* (2015).

[27] ESTE, A., GRINGOLI, F., AND SALGARELLI, L. Support vector machines for tcp traffic classification. *Computer Networks* (2009).

[28] FEURER, M., KLEIN, A., EGGENSPERGER, K., SPRINGENBERG, J. T., BLUM, M., AND HUTTER, F. Auto-sklearn: Efficient and Robust Automated Machine Learning. In *Automated Machine Learning*. 2019.

[29] FIRESTONE, D., PUTNAM, A., MUNDKUR, S., CHIOU, D., DABAGH, A., ANDREWARTHA, M., ANGEPAT, H., BHANU, V., CAULFIELD, A., CHUNG, E., CHANDRAPPA, H. K., CHATURMOHTA, S., HUMPHREY, M., LAVIER, J., LAM, N., LIU, F., OVTCHAROV, K., PADHYE, J., POPURI, G., RAINDEL, S., SAPRE, T., SHAW, M., SILVA, G., SIVAKUMAR, M., SRIVASTAVA, N., VERMA, A., ZUHAIR, Q., BANSAL, D., BURGER, D., VAID, K., MALTZ, D. A., AND GREENBERG, A. Azure Accelerated Networking: SmartNICs in the Public Cloud. In *USENIX NSDI* (2018).

[30] GARDNER, J. R., KUSNER, M. J., XU, Z. E., WEINBERGER, K. Q., AND CUNNINGHAM, J. P. Bayesian Optimization with Inequality Constraints. In *ICML* (2014).

[31] GELBART, M. A., SNOEK, J., AND ADAMS, R. P. Bayesian Optimization with Unknown Constraints. *arXiv preprint arXiv:1403.5607* (2014).

[32] GENG, Y., LIU, S., YIN, Z., NAIK, A., PRABHAKAR, B., ROSENBLUM, M., AND VAHDAT, A. SIMON: A Simple and Scalable Method for Sensing, Inference and Measurement in Data Center Networks. In *USENIX NSDI* (2019).

[33] HA, S., RHEE, I., AND XU, L. CUBIC: A New TCP-Friendly High-Speed TCP Variant. *ACM SIGOPS Operating Systems Review* (2008).

[34] HARIRI, B., AND SADATI, N. NN-RED: An AQM Mechanism Based on Neural Networks. *Electronics Letters* (2007).

[35] HE, K., ZHANG, X., REN, S., AND SUN, J. Deep Residual Learning for Image Recognition. In *IEEE CVPR* (2016).

[36] He, X., Zhao, K., and Chu, X. AutoML: A Survey of the State-of-the-Art. *Knowledge-Based Systems* (2021).

[37] Hearst, M., Dumais, S., Osuna, E., Platt, J., and Scholkopf, B. Support vector machines. *IEEE Intelligent Systems and their Applications* (1998).

[38] Holland, J., Schmitt, P., Feamster, N., and Mittal, P. New Directions in Automated Traffic Analysis. In *ACM SIGSAC CCS* (2021).

[39] Huang, P., Guo, C., Zhou, L., Lorch, J. R., Dang, Y., Chintalapati, M., and Yao, R. Gray Failure: The Achilles' Heel of Cloud-Scale Systems. In *HotOS* (2017).

[40] Hutter, F., Kotthoff, L., and Vanschoren, J. *Automated Machine Learning: Methods, Systems, Challenges*. Springer Nature, 2019.

[41] Ibanez, S., Brebner, G., McKeown, N., and Zilberman, N. The P4->NetFPGA Workflow for Line-Rate Packet Processing. In *ACM/SIGDA FPGA* (2019).

[42] Intel. Infrastructure Processing Unit (Intel IPU) and SmartNICs. https://www.intel.com/content/www/us/en/products/network-io/smartnic.html, last accessed: 06/10/2022.

[43] Intel. Intel P4 Studio. https://www.intel.com/content/www/us/en/products/network-io/programmable-ethernet-switch/p4-suite/p4-studio.html, last accessed: 06/10/2022.

[44] Intel. Tofino: P4-programmable Ethernet Switch ASIC that Delivers Better Performance at Lower Power. https://www.intel.com/content/www/us/en/products/network-io/programmable-ethernet-switch/tofino-series.html, last accessed: 06/10/2022.

[45] Intel. Tofino2: Second-generation P4-programmable Ethernet Switch ASIC that Continues to Deliver Programmability without Compromise. https://www.intel.com/content/www/us/en/products/network-io/programmable-ethernet-switch/tofino-2-series.html, last accessed: 06/10/2022.

[46] Jay, N., Rotman, N., Godfrey, B., Schapira, M., and Tamar, A. A Deep Reinforcement Learning Perspective on Internet Congestion Control. In *ICML* (2019).

[47] Jin, H., Song, Q., and Hu, X. Auto-Keras: An Efficient Neural Architecture Search System. In *ACM SIGKDD* (2019).

[48] Kamath, R., and Sivalingam, K. M. Machine learning based flow classification in dcns using p4 switches. In *2021 International Conference on Computer Communications and Networks (ICCCN)* (2021), IEEE, pp. 1–10.

[49] Katta, N., Hira, M., Kim, C., Sivaraman, A., and Rexford, J. HULA: Scalable Load Balancing Using Programmable Data Planes. In *ACM SOSR* (2016).

[50] Knudde, N., van der Herten, J., Dhaene, T., and Couckuyt, I. Gpflowopt: A bayesian optimization library using tensorflow. *arXiv preprint arXiv:1711.03845* (2017).

[51] Koeplinger, D., Feldman, M., Prabhakar, R., Zhang, Y., Hadjis, S., Fiszel, R., Zhao, T., Nardi, L., Pedram, A., Kozyrakis, C., and Olukotun, K. Spatial: A Language and Compiler for Application Accelerators. In *ACM SIGPLAN PLDI* (2018).

[52] Lapolli, �. C., Adilson Marques, J., and Gaspary, L. P. Offloading real-time ddos attack detection to programmable data planes. In *2019 IFIP/IEEE Symposium on Integrated Network and Service Management (IM)* (2019).

[53] Li, G., Zhang, M., Wang, S., Liu, C., Xu, M., Chen, A., Hu, H., Gu, G., Li, Q., and Wu, J. Enabling performant, flexible and cost-efficient ddos defense with programmable switches. *IEEE/ACM Transactions on Networking* (2021).

[54] Li, Y., Miao, R., Liu, H. H., Zhuang, Y., Feng, F., Tang, L., Cao, Z., Zhang, M., Kelly, F., Alizadeh, M., and Yu, M. HPCC: High Precision Congestion Control. In *ACM SIGCOMM* (2019).

[55] Liu, Y., Li, W., and Li, Y. Network Traffic Classification Using K-means Clustering. In *IMSCCS* (2007).

[56] Liu, Z., Namkung, H., Nikolaidis, G., Lee, J., Kim, C., Jin, X., Braverman, V., Yu, M., and Sekar, V. Jaqen: A High-Performance Switch-Native Approach for Detecting and Mitigating Volumetric DDoS Attacks with Programmable Switches. In *USENIX Security* (2021).

[57] Lockwood, J. W., McKeown, N., Watson, G., Gibb, G., Hartke, P., Naous, J., Raghuraman, R., and Luo, J. NetFPGA–An Open Platform for Gigabit-rate Network Switching and Routing. In *IEEE MSE* (2007).

[58] Mao, H., Alizadeh, M., Menache, I., and Kandula, S. Resource Management with Deep Reinforcement Learning. In *ACM HotNets* (2016).

[59] Mao, H., Netravali, R., and Alizadeh, M. Neural Adaptive Video Streaming with Pensieve. In *ACM SIGCOMM* (2017).

[60] Mao, H., Schwarzkopf, M., Venkatakrishnan, S. B., Meng, Z., and Alizadeh, M. Learning scheduling algorithms for data processing clusters. In *Proceedings of the ACM Special Interest Group on Data Communication* (2019).

[61] McKeown, N., Anderson, T., Balakrishnan, H., Parulkar, G., Peterson, L., Rexford, J., Shenker, S., and Turner, J. OpenFlow: Enabling Innovation in Campus Networks. *ACM SIGCOMM CCR 38*, 2 (2008), 69–74.

[62] McKinney, W., et al. pandas: A Foundational Python Library for Data Analysis and Statistics. *Python for High Performance and Scientific Computing 14*, 9 (2011), 1–9.

[63] Mehmood, T., and Rais, H. B. M. SVM for Network Anomaly Detection using ACO Feature Subset. In *IEEE iSMSC* (2015).

[64] Mockus, J., Tiesis, V., and Zilinskas, A. The Application of Bayesian Methods for Seeking the Extremum. *Toward Global Optimization 2*, 117-129 (1978), 2.

[65] Moody, J. Prediction Risk and Architecture Selection for Neural Networks. In *From Statistics to Neural Networks*. Springer, 1994, pp. 147–165.

[66] Narang, P., Ray, S., Hota, C., and Venkatakrishnan, V. Peershark: Detecting Peer-to-Peer Botnets by Tracking Conversations. In *IEEE Security and Privacy Workshop* (2014).

[67] Nardi, L., Bodin, B., Saeedi, S., Vespa, E., Davison, A. J., and Kelly, P. H. Algorithmic Performance-accuracy Trade-off in 3D Vision Applications using Hypermapper. In *IEEE IPDPSW* (2017).

[68] Nardi, L., Koeplinger, D., and Olukotun, K. Practical Design Space Exploration. In *IEEE MASCOTS* (2019).

[69] Nvidia. Bluefield Data Processing Units (DPUs). https://www.nvidia.com/en-us/networking/products/data-processing-unit/, last accessed: 06/10/2022.

[70] ONF. ONOS: Open Network Operating System (ONOS). https://opennetworking.org/onos/, last accessed: 06/10/2022.

[71] ONF. Stratum: Enabling the era of next generation SDN. https://opennetworking.org/stratum/, last accessed: 06/10/2022.

[72] Paria, B., Kandasamy, K., and Póczos, B. A Flexible Framework for Multi-Objective Bayesian Optimization using Random Scalarizations. In *UAI* (2019).

[73] Poupart, P., Chen, Z., Jaini, P., Fung, F., Susanto, H., Geng, Y., Chen, L., Chen, K., and Jin, H. Online Flow Size Prediction for Improved Network Routing. In *IEEE ICNP* (2016).

[74] Prabhakar, R., Zhang, Y., Koeplinger, D., Feldman, M., Zhao, T., Hadjis, S., Pedram, A., Kozyrakis, C., and Olukotun, K. Plasticine: A Reconfigurable Architecture for Parallel Patterns. In *ACM/IEEE ISCA* (2017).

[75] Putnam, A. FPGAs in the Datacenter: Combining the Worlds of Hardware and Software Development. In *GLSVLSI* (2017).

[76] Putnam, A., Caulfield, A. M., Chung, E. S., Chiou, D., Constantinides, K., Demme, J., Esmaeilzadeh, H., Fowers, J., Gopal, G. P., Gray, J., Haselman, M., Hauck, S., Heil, S., Hormati, A., Kim, J.-Y., Lanka, S., Larus, J., Peterson, E., Pope, S., Smith, A., Thong, J., Xiao, P. Y., and Burger, D. A Reconfigurable Fabric for Accelerating Large-scale Datacenter Services. In *ACM/IEEE ISCA* (2014).

[77] Rahbarinia, B., Perdisci, R., Lanzi, A., and Li, K. PeerRush: Mining for Unwanted P2P Traffic. In *DIMVA* (2013).

[78] Real, E., Moore, S., Selle, A., Saxena, S., Suematsu, Y. L., Tan, J., Le, Q. V., and Kurakin, A. Large-scale Evolution of Image Classifiers. In *ICML* (2017).

[79] Saeedi, S., Nardi, L., Johns, E., Bodin, B., Kelly, P. H., and Davison, A. J. Application-oriented Design Space Exploration for SLAM Algorithms. In *IEEE ICRA* (2017).

[80] Sanvito, D., Siracusano, G., and Bifulco, R. Can the Network Be the AI Accelerator? In *ACM NetCompute* (2018).

[81] Siracusano, G., and Bifulco, R. In-network Neural Networks. *arXiv preprint arXiv:1801.05731* (2018).

[82] Sonchack, J., Loehr, D., Rexford, J., and Walker, D. Lucid: A Language for Control in the Data Plane. In *ACM SIGCOMM* (2021).

[83] Song, L., Vempala, S., Wilmes, J., and Xie, B. On the Complexity of Learning Neural Networks. *arXiv preprint arXiv:1707.04615* (2017).

[84] Souza, A., Nardi, L., Oliveira, L. B., Olukotun, K., Lindauer, M., and Hutter, F. Bayesian Optimization with a Prior for the Optimum. In *Joint European Conference on Machine Learning and Knowledge Discovery in Databases* (2021).

[85] Swamy, T., Rucker, A., Shahbaz, M., Gaur, I., and Olukotun, K. Taurus: A Data Plane Architecture for per-Packet ML. In *ASPLOS* (2022).

[86] Tang, T. A., Mhamdi, L., McLernon, D., Zaidi, S. A. R., and Ghogho, M. Deep Learning Approach for Network Intrusion Detection in Software Defined Networking. In *IEEE WINCOM* (2016).

[87] Tavallaee, M., Bagheri, E., Lu, W., and Ghorbani, A. A. A Detailed Analysis of the KDD CUP 99 Dataset. In *IEEE CISDA* (2009).

[88] Torrey, L., and Shavlik, J. Transfer Learning. In *Handbook of Research on Machine Learning Applications and Trends: Algorithms, Methods, and Techniques*. IGI global, 2010, pp. 242–264.

[89] Đukić, V., Jyothi, S. A., Karlas, B., Owaida, M., Zhang, C., and Singla, A. Is Advance Knowledge of Flow Sizes a Plausible Assumption? In *USENIX NSDI* (2019).

[90] Weiss, K., Khoshgoftaar, T. M., and Wang, D. A Survey of Transfer Learning. *Journal of Big data 3*, 1 (2016), 1–40.

[91] Winstein, K., and Balakrishnan, H. TCP ex machina: Computer-generated Congestion Control. In *ACM SIGCOMM CCR* (2013).

[92] Xilinx. Vivado. https://www.xilinx.com/products/design-tools/vivado.html.

[93] Xilinx. Alveo U250 Data Center Accelerator Card. https://www.xilinx.com/products/boards-and-kits/alveo/u250.html, last accessed: 06/10/2022.

[94] Xilinx. Running Multiple Implementation Strategies for Timing Closure. https://docs.xilinx.com/r/en-US/ug1393-vitis-application-acceleration/Running-Multiple-Implementation-Strategies-for-Timing-Closure, last accessed: 06/10/2022.

[95] Xilinx. Ultrascale+ integrated 100g ethernet subsystem. https://www.xilinx.com/products/intellectual-property/cmac_usplus.html, last accessed: 06/10/2022.

[96] Xiong, Z., and Zilberman, N. Do Switches Dream of Machine Learning? Toward In-network Classification. In *ACM HotNets* (2019).

[97] Yan, F. Y., Ayers, H., Zhu, C., Fouladi, S., Hong, J., Zhang, K., Levis, P., and Winstein, K. Learning in situ: A Randomized Experiment in Video Streaming. In *USENIX NSDI* (2020).

[98] Yan, F. Y., Ma, J., Hill, G. D., Raghavan, D., Wahby, R. S., Levis, P., and Winstein, K. Pantheon: The Training Ground for Internet Congestion-Control Research. In *USENIX ATC* (2018).

[99] Yan, S., Wang, X., Zheng, X., Xia, Y., Liu, D., and Deng, W. ACC: Automatic ECN Tuning for High-Speed Datacenter Networks. In *ACM SIGCOMM* (2021).

[100] Yu, L., Sonchack, J., and Liu, V. Mantis: Reactive Programmable Switches. In *ACM SIGCOMM* (2020).

[101] Zhang, Q., Liu, V., Zeng, H., and Krishnamurthy, A. High-Resolution Measurement of Data Center Microbursts. In *ACM IMC* (2017).

[102] Zhang, Y., Zhang, N., Zhao, T., Vilim, M., Shahbaz, M., and Olukotun, K. SARA: Scaling a Reconfigurable Dataflow Accelerator. In *ACM/IEEE ISCA* (2021).

[103] Zhu, Y., Eran, H., Firestone, D., Guo, C., Lipshteyn, M., Liron, Y., Padhye, J., Raindel, S., Yahia, M. H., and Zhang, M. Congestion Control for Large-Scale RDMA Deployments. In *ACM SIGCOMM* (2015).

[104] Zilberman, N., Audzevich, Y., Covington, G. A., and Moore, A. W. NetFPGA SUME: Toward 100 Gbps as Research Commodity. *IEEE Micro 34*, 5 (2014), 32–41.

[105] Zimmer, L., Lindauer, M., and Hutter, F. Auto-Pytorch: Multi-Fidelity MetaLearning for Efficient and Robust AutoDL. *IEEE Transactions on Pattern Analysis and Machine Intelligence* (2021).